\documentclass[11pt,a4paper]{article}

\usepackage{authblk}

\usepackage[english]{babel}
\usepackage[latin1]{inputenc}
\usepackage[T1]{fontenc}

\usepackage[top=26mm, bottom=26mm, left=24mm, right=24mm]{geometry}

\usepackage{color}

\usepackage{amsfonts}
\usepackage{amsmath}
\usepackage{amsthm}
\usepackage{bm}
\DeclareMathAlphabet{\mathpzc}{OT1}{pzc}{m}{it}

\usepackage{authblk}

\theoremstyle{definition}

\theoremstyle{remark}

\newcommand{\be}{\begin{equation}}
\newcommand{\ee}{\end{equation}}
\newcommand{\bea}{\begin{eqnarray}}
\newcommand{\eea}{\end{eqnarray}}
\newcommand{\nn}{\nonumber}

\newcommand{\DIV}[1][\normalsize]{\,\mbox{div}\,}

\newcommand{\res}{\mathrm{res}}
\newcommand{\tr}{\mathrm{tr}}

\def\N{ {\mathbb N} }

\def\R{ {\mathbb R} }

\def\dude1#1{{\color [rgb]{0,0.6,0.6} [some dude: #1]}}
\def\dude1#1{{\color [rgb]{0.6,0.0,0.6} [some other dude: #1]}}

\begin{document}

\title{Modified heat equations for an analytic continuation of the spectral $\zeta$ function}

\author{Tobias Zingg \thanks{zingg@nordita.org}}

\affil{Nordita, Stockholm University and KTH Royal Institute of Technology,
Roslagstullsbacken 23, SE-106 91 Stockholm, Sweden}


\date{\today}

\maketitle
\begin{center}
\begin{abstract}

%

For an elliptic differential operator $D$ of order $h$ in $n$ dimensions, the spectral $\zeta$-function $\zeta_D(s)$ for $\Re s > \frac{n}{h}$ can be evaluated as an integral over the heat kernel $e^{-t D}$.
Here, alternative expressions for $\zeta_D(s)$ are presented involving an integral over kernels $\mathpzc{k}_{n,m}$ for a modified heat equation, such that the integral is non-singular around $s=0$, respectively close to potential poles around $s=\frac{m}{h}, m<n$.
Besides explicit expressions for an analytic continuation of $\zeta_D(s)$ when $\Re s \leq \frac{n}{h}$, this provides an alternative method to study functional determinants and the residues of $\zeta_D(s)$ that does not require to compute Seeley--DeWitt coefficients explicitly to cancel divergences in the heat trace.

\end{abstract}
\end{center}

\thispagestyle{empty}
\tableofcontents

\newpage
\renewcommand{\thefootnote}{\arabic{footnote}}
\setcounter{footnote}{0}
\setcounter{page}{1}

%

\newpage

\section{Introduction}
\label{sec:intro}

The $\zeta$ function for elliptic operators finds many applications in mathematics and theoretical physics. It is associated with the spectrum of the operator, and in particular for Laplace and Dirac type operators on curved manifolds it also encodes information about the geometry -- see e.g.~\cite{Gilkey:1995mj}.
As such, $\zeta$ function regularization~\cite{Dowker:1975tf,Hawking:1976ja} has become an invaluable tool in many applications for quantum field theory, especially on curved space-times, involving vacuum polarization and the Casimir effect, the study of quantum anomalies, as well as one-loop calculations in general -- see e.g.~\cite{Birrell:1982ix,elizalde1995ten} for a comprehensive overview.

To recap the basic concepts, consider $D$, an elliptic positive definite operator of order $h$ on a compact manifold of dimension $n$. If $\{\lambda_n,\phi_n\}_{n\in\N}$ is the spectral decomposition, then the spectral $\zeta$ function is
\bea
\zeta_D(s)
	&:=&	\sum_{n\geq 0} \lambda_n^{-s}	\, ,
\label{def:zeta_EW}
\eea
for $s$ large enough such that the sum converges. This can then be extended to a meromorphic function. In particular, the $\zeta$- regularized determinant~\cite{RAY1971145} is given by\footnote{up to a choice of renormalization scale that will not be relevant in what follows and thus ignored}
\bea
\ln {\det}_\zeta D &:=& -\zeta'_D(0)
\, .
\label{eq:func_det}
\eea
To evaluate and analyze $\zeta_D(s)$ it has been found to be quite effective to use the heat kernel~\cite{Heat_kernels_and_Dirac_operators,Vassilevich:2003xt}. For $\Re s> n/h$,
\bea
\zeta_D(s)
	&=&	\frac{1}{\Gamma(s)} \int_{0}^{\infty} t^{s-1} K(t,D) \, dt
\label{def:zeta_int}
\eea
with the heat kernel trace
\bea
K(t,D)	&=&	\mathrm{tr} \, e^{-tD}	\, ,
\label{def:heat_trace}
\eea
where $e^{-tD}$ is interpreted as the functional that maps a function $\psi_0$ to a solution of the initial value problem
\bea
\left\{
\begin{array}{l}
\partial_t \Psi + D \Psi = 0	\, ,	\\
\Psi(0) = \psi_0	\, .
\end{array}\right.
\label{eq:heat_ivp}
\eea
What makes~\eqref{def:zeta_int} a useful expression to study properties of $\zeta_D(s)$ is that the heat kernel trace~\eqref{def:heat_trace} has an asymptotic expansion~\cite{Minakshisundaram:1949xg,Greiner:1971xe} that is rather well understood,
\bea
K(t,D)	&\sim &	\sum_{n>0} a_k(D) t^{-\frac{n-k}{h}}	\, .
\label{eq:heat_kernel_expansion}
\eea
By computing the $a_k(D)$, which are related to geometric invariants and boundary conditions, it is possible to analytically continue~\eqref{def:zeta_int} to values with $\Re s \leq n/h$. And it is found that the only possible singularities of ${\zeta_D}(s)$ are simple poles which can only be located at $s=(n-k)/h$, $k\in\N$. The residues at these poles are $a_k(D)/\Gamma(\frac{n-k}{h})$.

Despite many results being known -- in particular for $n=1$ or standard Laplacian or Dirac type operators~\cite{Heat_kernels_and_Dirac_operators,Gelfand:1959nq,d'hoker1986,Shimon:1977,Burghelea:1995} -- it can still a rather challenging and non-trivial task to actually compute ${\zeta_D}(s)$ or $a_k(D)$ for specific values or more generic operators.
The motivation of this paper is therefore to provide an alternative, complementary route to analytically continue ${\zeta_D}(s)$. This is achieved by using knowledge of the asymptotic expansion~\eqref{eq:heat_kernel_expansion} to modify the integrand in~\eqref{def:zeta_int} such that the integral -- apart from the isolated poles, of course -- is convergent for values with $\Re s \geq \frac{n-m}{h}$. These integrands are then related to initial value problems involving a modified heat equation.

The paper is organized as follows. In section~\ref{sec:s0} it is demonstrated how an analytic continuation of~\eqref{def:zeta_int} to $s=0$ can be obtained by modifying the integrand. In section~\ref{sec:residues}, the same ideas are used to derive alternative expressions to continue $\zeta_D(s)$ to $\Re s \geq \frac{n-m}{h}$. These continuations provide alternative ways to express the $\zeta$ regularized determinant and the residues that do not involve an explicit computation of the coefficients in the heat kernel expansion, but instead rely on studying initial value problems for a modified heat equation.
Section~\ref{sec:psidiff} provides some considerations on how these ideas could be extended to the case of pseudo-differential operators with non-integer order $h$.

\section{Analytic continuation to $s=0$}
\label{sec:s0}

It is well-known that $\zeta_D(s)$, when analytically extended into a meromorphic function, is regular at $s=0$. As mentioned above, due to the divergent terms in~\eqref{eq:heat_kernel_expansion}, the integral~\eqref{def:zeta_int} is not convergent for $\Re s > \frac{n}{h}$. A common method to evaluate $\zeta'_D(0)$ is to calculate the coeffcients $a_k(D)$ for the divergent terms -- which, in principle, can be worked out as functionals of curvature and boundary conditions -- and subtract them from the heat trace.

Here, an alternative way is presented, in which the integrand is changed from the common heat trace to a kernel for a modified heat equation.
For this purpose, change variable~$t\to\tau^h$ such that,
\bea
\zeta_D(s)
	&=&	\frac{h}{\Gamma(s)} \int_0^\infty \tau^{hs-1-n} \,\tau^n K(\tau^h,D) \, d\tau	\, .
\label{eq:zeta_ttotau}
\eea
With the asymptotic expansion~\eqref{eq:heat_kernel_expansion} and the assumption that manifold and boundary conditions are sufficiently smooth, $\tau^n K(\tau^h,D)$ is regular at $\tau=0$ by construction and falls of exponentially for $\tau\to\infty$. Thus, after integrating by parts $n+1$ times,
\bea
{\zeta_D}(s)
	&=&	(-1)^{n} \frac{h \Gamma(hs-n)}{\Gamma(s)\Gamma(hs+1)} \int_0^\infty \tau^{hs} \mathpzc{K}_n(\tau,D)\, d\tau
\, ,
\label{eq:zeta_KtocurlyK}
\eea
where the kernel
\bea
\mathpzc{K}_n(\tau,D)
	&=&	- \frac{\partial^{n+1}}{\partial\tau^{n+1}} \tau^n K(\tau^n,D)	\, ,
\label{def:curlyK}
\eea
has been introduced. With the integrand being regular for $\Re s> -1/h$, the expression in~\eqref{eq:zeta_KtocurlyK} becomes an analytic continuation for  $\zeta_D(s)$.\footnote{That is evaluates to $\zeta_D(s)$ for $\Re s> n/h$ can easily be checked via a spectral decomposition.}
In particular, close to $s=0$, after rearranging the terms in the prefactor,
\bea
{\zeta_D}(s)
	&=&	\frac{\Gamma(1-hs)}{\Gamma(s+1)\Gamma(n+1-hs)} \int_0^\infty \tau^{hs} \mathpzc{K}_n(\tau,D) \, d\tau	\, .
\label{eq:zeta_KtocurlyK_s0}
\eea
Now, it is straightforward to evaluate,
\bea
{\zeta_D}(0)	&=&	\frac{1}{\Gamma(n+1)} \int_0^\infty \mathpzc{K}_n(\tau,D) \, d\tau	\, ,		\\
{\zeta'_D}(0)	&=&	\frac{1}{\Gamma(n+1)} \int_0^\infty \left( \gamma+h H_n + h \ln\tau \right) \mathpzc{K}_n(\tau,D) \, d\tau	\, .
\label{eq:zeta0}
\eea
The rest of the section will deal with characterizing $\mathpzc{K}_n(\tau,D)$ as a kernel trace that is related to an initial value problem, similarly as $e^{-tD}$ can be identified as the kernel related to~\eqref{eq:heat_ivp}.

\subsection{$\mathpzc{K}_n(\tau,D)$ as kernel trace for a modified heat equation}

Firstly, rewrite~\eqref{def:curlyK} as 
\bea
\mathpzc{K}_n(\tau,D)
	&=&	\tr\, \mathpzc{k}_n(\tau,D)	\, ,
\label{def:curlyK_alt}
\eea
which introduces the functional
\bea
\mathpzc{k}_n(\tau,D)
	&=&	- \frac{\partial^{n+1}}{\partial\tau^{n+1}} \tau^n e^{\tau^h D}	\, .
\label{def:curlyk}
\eea
Then, expanding the exponential into a series,
\bea
\mathpzc{k}_n
	\;=\;	-\partial_\tau^{n+1} \sum_{k \geq 0} \frac{ \tau^{n+hk} (-D)^k }{\Gamma(k+1)}	
	\;=\;	\tau^{h-1} D \sum_{k \geq 0} \frac{\Gamma(hk+h+n+1)}{\Gamma(k+2) \Gamma(hk+h)} \left(-\tau^h D\right)^k
	\, .
\label{eq:curlykntoF_1}
\eea
Using the multiplication formula for the $\Gamma$ function,
\bea
\Gamma(hz+b)
	&=&	h^{hz+b-1/2} (2\pi)^{\frac{1-n}{2}} \prod_{l=0}^{h-1} \Gamma\left( z + \frac{b+l}{h} \right)	\, ,
\label{eq:Gamma_mult}
\eea
it is possible to rewrite~\eqref{eq:curlykntoF_1} as
\bea
\mathpzc{k}_n
	&=&	\frac{\Gamma(h+n+1)}{\Gamma(h)} \tau^{h-1} D \sum_{k \geq 0} \frac{ \left(-\tau^h D\right)^k }{ \Gamma(k+2) } \prod_{l=0}^{h-1}  \frac{ \Gamma\left( k+1+\frac{n+l+1}{h} \right) \Gamma\left( 1+\frac{l}{h} \right) }{ \Gamma\left( 1+\frac{n+l+1}{h} \right) \Gamma\left( k+1+\frac{l}{h} \right) } \nn\\
	&=&	\frac{\Gamma(h+n+1)}{\Gamma(h)} \tau^{h-1} D\, {_h F_h}\left[ 1+\frac{n+1}{h}, \ldots , 1+\frac{n+h+1}{h}; 1+\frac{1}{h}, \ldots , 2 ; -\tau^h D \right]	\, .	\quad
\label{eq:curlykntoF_2}
\eea
With ${_h F_h}$ being a solution to a hypergeometric equation of order $h+1$, it follows that $\Psi=\mathpzc{k}_n \psi_0$ can be characterized as the unique regular solution to an initial value problem as well. Specifically, For the common cases $h=1,2$,
\bea
h=1 &:&
\left\{
\begin{array}{l}
\left[ \partial_\tau \left(\tau \partial_\tau + 1 \right) + \left(\tau \partial_\tau + n+2 \right) D \right]\Psi = 0	\, ,	\\
\Psi(0) = \Gamma(n+2)D\psi_0		\, ,	\\
\dot{\Psi}(0) = -\frac{\Gamma(n+3)}{6}D^2\psi_0	\, .
\end{array}\right.	\\
h=2 &:&
\left\{
\begin{array}{l}
\left[ \partial_\tau \left(\tau \partial_\tau +1 \right)\left(\tau \partial_\tau - 1 \right) + 2\tau\left(\tau \partial_\tau + n+2 \right)\left(\tau \partial_\tau + n+3 \right) D \right]\Psi = 0	\, ,	\\
\Psi(0) = 0		\, ,	\\
\dot{\Psi}(0) = \Gamma(n+h+1)D\psi_0	\, ,	\\
\ddot{\Psi}(0) = 0		\, .
\end{array}\right.\qquad
\eea
%
%
And generally, for $h>1$,
\bea
\left\{
\begin{array}{l}
\left[ \frac{1}{\tau}\prod\limits_{j=0}^h \left(\tau \partial_\tau +j-h+1 \right) + h\tau^{h-1} D \prod\limits_{j=1}^h \left(\tau \partial_\tau +j+n+1 \right) \right]\Psi = 0	\, ,	\\
\Psi^{(j)}(0) = 0		\, , \, 0\leq j < h-1	\, ,\\
\Psi^{(h-1)}(0) = \Gamma(n+h+1)D\psi_0	\, ,	\\
\Psi^{(h)}(0) = 0		\, .
\end{array}\right.
\label{eq:s0_ivp}
\eea

\section{Residues}
\label{sec:residues}

Analogously to the previous section~\ref{sec:s0}, it is also possible to modify the integrand in~\eqref{def:zeta_int} such that an evaluation around one of the potential poles at $s=-\frac{m}{h}, m\geq-n$ becomes more convenient.
Again, from~\eqref{eq:zeta_ttotau}, but this time integrating $n+m$ times by parts,
\bea
{\zeta_D}(s)
	&=&	(-1)^{n+m} \frac{h \Gamma(hs-n)}{\Gamma(s)\Gamma(hs+m+1)} \int_0^\infty \tau^{hs+m} \mathpzc{K}_{n,m}(\tau,D)\, d\tau
\, ,
\label{eq:zeta_KtocurlyKnm}
\eea
which introduced
\bea
\mathpzc{K}_{n,m}(\tau,D)
	\;=\;	- \frac{\partial^{n+m+1}}{\partial\tau^{n+m+1}} \tau^n K(\tau^n,D)
	\;=\;	\tr \mathpzc{k}_{n,m}(\tau,D)	\, ,
\label{def:curlyKmn}
\eea
with the kernels
\bea
\mathpzc{k}_{n,m}(\tau,D)
	&=&	- \frac{\partial^{n+m+1}}{\partial\tau^{n+m+1}} \tau^n e^{\tau^h D}	\, .
\label{def:curlyknm}
\eea
Considering that
\bea
\frac{\Gamma(hs-n)}{\Gamma(s)}
	&=&	\frac{\Gamma(1-s) \sin(\pi s)}{\Gamma(1+n-hs) \sin(\pi h s - \pi n)}
\, ,
\eea
the residue of~\eqref{eq:zeta_KtocurlyKnm} at $s=-\frac{m}{h}$ turns out to be
\bea
\res_{s=-\frac{m}{h}} \zeta_D(s)
	\;=\;	- \frac{\sin\left( \frac{\pi m}{h} \right) \Gamma\left( \frac{h + m}{h} \right)}{\pi \Gamma\left( 1+n+m \right)} \int\limits_0^\infty \mathpzc{K}_{n,m}(\tau,D)\, d\tau
\;=\;	\frac{\sin\left( \frac{\pi m}{h} \right) \Gamma\left( \frac{h + m}{h} \right)}{\pi \Gamma\left( 1+n+m \right)} \mathpzc{K}_{n,m-1}(0,D)
\, .\;
\label{eq:residues}
\eea
This also makes manifest that there are no poles when $m \in h\N$. For positive values of $s$ it might be more convenient to write this as
\bea
\res_{s=\frac{\mu}{h}} \zeta_D(s)
	\;=\;	\frac{1}{\Gamma\left( \frac{\mu}{h} \right) \Gamma\left( 1+n-\mu \right)} \int\limits_0^\infty \mathpzc{K}_{n,-\mu}(\tau,D)\, d\tau
\, .\;
\label{eq:residues_sneg}
\eea
As before, the kernels~\eqref{def:curlyknm} can be associated with modified heat equations, which is worked out in the subsequent sections.

\subsection{Residues with $s<0$}
\label{ses:sneq}

Write $m=qh+r$ with $q,r \in \N$ and $0 \leq r <h$. After expanding~\eqref{def:curlyknm} into a power series,
\bea
\mathpzc{k}_{n,m}
	&=&	- \sum_{k > q} \frac{\Gamma(hk+n+1)}{\Gamma(k+1) \Gamma(hk-qh-r)} \tau^{hk-qh-r-1} \left(- D\right)^k	 \nn\\
	&=&	- \tau^{h-r-1} (-D)^{q+1} \sum_{k \geq 0} \frac{\Gamma(hk+qh+h+n+1)}{\Gamma(k+q+2) \Gamma(hk-qh-r)} \left(- \tau^h D\right)^k
\, .	\quad
\label{eq:curlyknmtoF_sneg_1}
\eea
Using again the multiplication formula~\eqref{eq:Gamma_mult} to rewrite the term in the sum, this expression can be expressed as a hypergeometric function, 
\bea
\mathpzc{k}_{n,m}
	&=&	- \tau^{h-r-1} (-D)^{q+1} \frac{\Gamma(hq+h+n+1)}{\Gamma(q+2)\Gamma(h-r)} \, {_h F_h}\left[ \{ a_l \} ; \{b_l\} ; -\tau^h D \right]	\, ,	\quad
\label{eq:curlyknmtoF_sneg_2}
\eea
with the coefficients
\bea
\left\{ a_l \right\}_{l=1}^{h}	&=&	\left\{ q+1+\frac{n+l}{h} \right\}_{l=1}^{h}	\, ,	\\
\left\{ b_l \right\}_{l=1}^{h}	&=&	\left\{ 1-\frac{r}{h}, \ldots , 1-\frac{1}{h}, 1+\frac{1}{h} \ldots , 1+\frac{h-1-r}{h}, q+2 \right\}	\, .
\eea
Thus, $\mathpzc{k}_{n,m}$ can be identified as the functional that maps a function $\psi_0$ to the unique regular solution for the initial value problem,\footnote{The special case $h=1$ is not listed here, as the residue~\eqref{eq:residues} would vanish.}
\bea
\left\{
\begin{array}{l}
\left[ \frac{1}{\tau}\left(\tau \partial_\tau + m \right)\prod\limits_{\stackrel{j=0}{j\neq r}}^h \left(\tau \partial_\tau - j \right) + h\tau^{h-1} D \prod\limits_{j=1}^h \left(\tau \partial_\tau +j+n+m+1 \right) \right]\Psi = 0	\, ,	\\
\Psi^{(j)}(0) = 0		\, , \; j \neq h-1-r \, , \; 0\leq j \leq h	\, ,\\
\Psi^{(h-r-1)}(0) = - \frac{\Gamma(hq+h+n+1)}{\Gamma(q+2)} (-D)^{q+1}\psi_0	\, .
\end{array}\right.
\label{eq:s0_ivp_spos}
\eea
In the case $m,q,r \to 0$, this becomes~\eqref{eq:s0_ivp}, as is to be expected.

\subsection{Residues with $s>0$}

For residues with positive values of $s$, set $m=-\mu$ in the following. Then, expanding~\eqref{def:curlyknm} into a series,
\bea
\mathpzc{k}_{n,m}
	&=&	- \sum_{k \geq 0} \frac{\Gamma(hk+n+1)}{\Gamma(k+1) \Gamma(hk+\mu)} \tau^{hk+\mu -1} \left(- D\right)^k	 \nn\\
	&=&	-\frac{\Gamma(n+1)}{\Gamma(\mu)} \tau^{\mu-1} \sum_{k \geq 0} \frac{ \left(-\tau^h D\right)^k }{ \Gamma(k+2) } \prod_{l=0}^{h-1}  \frac{ \Gamma\left( k+\frac{n+l+1}{h} \right) \Gamma\left( \frac{\mu+l}{h} \right) }{ \Gamma\left( \frac{n+l+1}{h} \right) \Gamma\left( k+\frac{\mu+l}{h} \right) } \nn\\
	&=&	- \frac{\Gamma(n+1)}{\Gamma(\mu)} \tau^{\mu-1} \, {_h F_h}\left[ \frac{n+1}{h}, \ldots , \frac{n+h}{h}; \frac{\mu}{h}, \ldots , \frac{\mu+h-1}{h} ; -\tau^h D \right]	\, .	\quad
\label{eq:curlyknmtoF}
\eea
In this case, $\Psi=\mathpzc{k}_{n,m}\psi_0$ solves the modified heat equation,
\bea
\left[\frac{1}{\tau} \left(\tau \partial_\tau - \mu +1 \right)\prod\limits_{j=1}^h \left(\tau \partial_\tau + j - h \right) + h\tau^{h-1} D \prod\limits_{j=1}^h \left(\tau \partial_\tau +j+n-\mu+1 \right) \right]\Psi = 0	\, .
\label{eq:modheat_spos}
\eea
Though, the situation with initial conditions is somewhat different from the previous cases. When $m>h$, initial conditions for the regular solution are not solution are not uniquely determined by the first $h$ derivatives - and this also includes one of the cases most relevant in many applications, which is a Laplacian on a manifold with dimension bigger than $2$.
In this situation, $\mathpzc{k}_{n,m}$ is characterized as the functional that maps to solutions of~\eqref{eq:modheat_spos} such that for small times,
\bea
\Psi	&\sim &	- \frac{\Gamma(n+1)}{\Gamma(\mu)} \tau^{\mu-1} + \mathpzc{o}\left(\tau^{\mu-1}\right)	\, .
\eea
This also covers the special case $h=1$, where there would be two regular linearly independent solutions.

\section{A note on pseudo-differential operators}
\label{sec:psidiff}

The previous section made use of $h\in \N$ to derive the modified heat equations related to the kernels $\mathpzc{k}_{n}$, respectively $\mathpzc{k}_{n,m}$. In principle, these kernels could also be defined for non-integer values of $h$, i.e.~in the case of pseudo-differential operators of that order. However, the respective initial value problems that characterize them would not any more be PDEs involving polynomial expressions in derivatives. Nevertheless, provided the heat kernel still has the asymptotic expansion~\eqref{eq:heat_kernel_expansion}, it is, in principle, possible to write some explicit expressions for integro-differential equations related to evaluating $\zeta_D(s)$ at specific points.
What is essentially needed is a mapping
\bea
\tau^a	&\to &	\frac{\Gamma(a)}{\Gamma(a+h)}	\, .
\eea
Thus, for a regular function $f$ consider,
\bea
\Xi_b[f](\tau)	&:=&	\int_0^1 f(\tau x) \left( 1-x \right)^{b-1}
\, .
\eea
By analytic continuation\footnote{e.g.~by changing the integration to a Pochhammer contour when required}, this can also be extended to functions which differ from a function regular on $\R_{\geq 0}$ by a generalized Laurent expansion around $\tau = 0$.
It is now a straightforward exercise to apply this term by term to the Frobenius expansion for $\Psi = \mathpzc{k}_{n,m}\psi_0$ to find that it solves the modified heat equation,
\bea
\partial_\tau \tau^{h-n} \Xi_h\left[\tau^{n+m-h+1}\Psi\right]
 + h \tau^{m+h} D \Xi_h\left[\Psi\right]	&=&	0
\, .
\label{eq:modheat_psidiff}
\eea
Further research would however be required to determine what types of initial conditions would be necessary to turn this into a well-posed initial value problem for generic values of $h$ and $m$.

\section{Conclusions}

A proposition was made that relates analytic continuation and residues of the $\zeta$ function for an elliptic operator $D$, as well as its $\zeta$ regulated functional determinant, to modified heat equations of the form
\bea
p(\tau,\partial_\tau) \Psi + D q(\tau,\partial_\tau)	&=& 0
\, ,
\eea
with polynomial expressions $p$ and $q$. The principal symbol of the operator in the PDE above, 
\bea
\tau^h \left[ \lambda + h\tau^{h-1} \sigma_D(\xi) \right] \lambda^h
\label{eq:principal_symbol}
\eea
is closely related to the symbol of heat equation, $\lambda + \sigma_D(\xi)$, and raises the question on what similar properties solutions will share with solutions to the heat equation -- e.g.~maximum principles and fall-off conditions. This will be left for future research.

One potential advantage of using these modified heat equations comes in numerical evaluations. Using the heat kernel requires to cancel divergences via calculating and subtracting Seeley--DeWitt coefficients. Though they are in principle known, formally at least, when it comes to explicit computation or numerical approximation it can still be a quite delicate process to evaluate these coefficients with sufficient precision to cancel divergences with the necessary numerical precision -- especially in applications like general relativity and back-reaction problems, where the geometry itself is a dynamical object. The kernels $\mathpzc{k}_{n,m}$ on the other hand do not require cancellation of divergences and are related to a modification of the heat equation that is, as far as numerical approximation is concerned, not much more complicated to deal with than the heat equation itself. Additionally, studying the $\mathpzc{k}_{n,m}$, respectively the related initial value problems, can also provide a new, alternative, way to calculate the Seeley--DeWitt coefficients. As the latter are not only appear in the heat kernel expansion but also in the residues of $\zeta_D
(s)$, it is in principle possible to evaluate the coefficients $a_k(D)$ by means of~(\ref{eq:residues},\ref{eq:residues_sneg}).

As a final remark, it is straightforward to generalize the results to the situation with operator insertions into the $\zeta$ function. Essentially, this simply requires to act with the operator on the initial conditions in the respective initial value problems for the modified heat equation derived above. This then also trivially extends to the $\eta$ invariant.



\bibliography{z-modHE}
\bibliographystyle{unsrt}

\end{document}